\documentstyle[prl,aps,epsfig]{revtex}

 \twocolumn
 \narrowtext

\begin{document}
\draft
\wideabs{
\title{Single Atoms in an Optical Dipole Trap:\\
Towards a Deterministic Source of Cold Atoms}
\author{D. Frese, B. Ueberholz, S. Kuhr, W. Alt, D. Schrader, V. Gomer and D. Meschede}
\address{Institut f\"ur Angewandte Physik, Universit\"at Bonn \\
         Wegelerstr.\ 8, D-53115 Bonn, Germany}
\maketitle
\begin{abstract}
 We describe a simple experimental technique which allows us to store a
 small and deterministic number of neutral atoms in an optical dipole
 trap. The desired atom number is prepared in a magneto-optical trap
 overlapped with a single focused Nd:YAG laser beam. Dipole trap loading
 efficiency of 100 \% and storage times of about one minute have been
 achieved. We have also prepared atoms in a certain hyperfine state and
 demonstrated the feasibility of a state-selective detection via resonance
 fluorescence at the level of a few neutral atoms. A spin relaxation time
 of the polarized sample of $4.2\pm 0.7$ s has been measured. Possible
 applications are briefly discussed.
\end{abstract}

 }

 Neutral atoms can conveniently and at very low kinetic energy be stored in
 a magneto-optical trap (MOT) \cite{Raab 87}, not only in large quantities
 but also in small and exactly known numbers of up to 20 single atoms
 \cite{Hu94,Haubrich 96}. For several applications, for instance in cavity
 quantum electrodynamics \cite{Berman 94}, it is of interest to use perfectly
 controlled or deterministic samples of atoms for further experiments
 involving quantum interactions of an exactly known number of atoms. Full
 control of all internal and external atomic degrees of freedom is necessary
 in such applications, but cannot be achieved in a MOT since due to its dissipative
 character all degrees of freedom are intimately mixed. In order to overcome
 this problem one can therefore combine the operating convenience of the MOT
 for isolated atoms with the advantages for quantum manipulation offered by
 the nearly conservative potential of optical dipole traps. The interest in
 optical dipole traps \cite{Gordon 80} as an elegant and simple way to store
 laser-cooled neutral atoms has rapidly increased within the last few years
 \cite{Grimm 99}. Far-off-resonance optical dipole traps \cite{Rolston 92} can
 confine atoms in all ground states for a long time with a very small
 ground-state relaxation rate \cite{Cline 94}.
 Cold atoms can be localized in micropotentials of a 3D periodic
 potential created by interfering laser beams (so-called optical lattices
 \cite{lattices}). Such experiments always operate with at least several
 thousands of atoms, the atom number can not be determined exactly
 and is controlled only on average.

 In the present work we load a small and exactly known atom number into an optical
 dipole trap with 100\% efficiency, opening up a route to a novel kind of cold atom
 sources free of the indeterminism intrinsic to usual sources like atom beams.
 We have also demonstrated the feasibility of a state-selective detection at the
 level of a few neutral atoms and measured a long spin relaxation time of some
 seconds. Together with recently demonstrated Raman sideband cooling
 \cite{Hamman 98} and the generation of non-classical motional states of atoms
 in standing-wave dipole traps \cite{Bouchoule 99}, this system promises to be a
 new basis for future experiments with full control of all atomic degrees of freedom.
 One of the most interesting possibilities would be
 long-time localization of more than one atom within a mode of a high finesse cavity.

\begin{figure}[hbt]
 \center{\epsfig{file=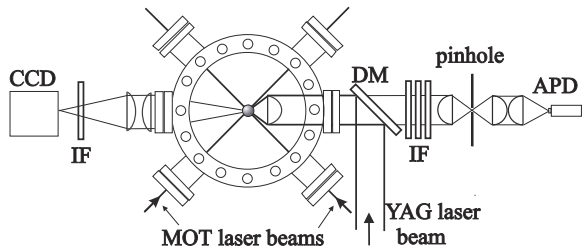,width=3in}}\vspace*{.4cm}
 \caption{
 Experimental setup. DM - dichroic mirror, IF - interference filters.
 In the image plane of the telescope a 150 $\mu$m pinhole is placed
 for spatial filtering of the stray light.
 }
 \label{setup}
 \end{figure}

\begin{figure}[hbt]
 \center{\epsfig{file=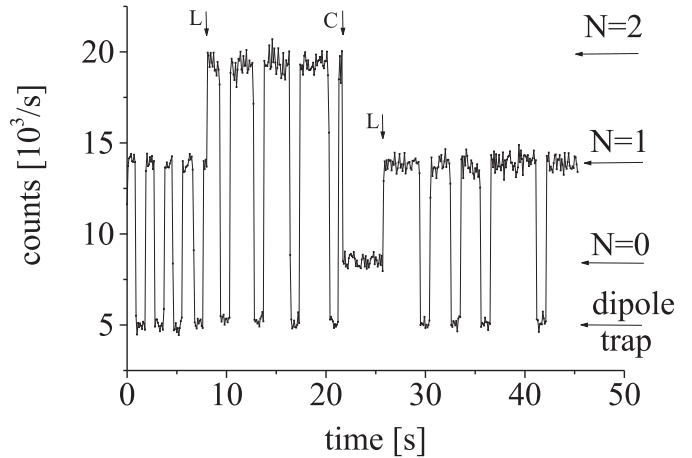,width=3.5in}}\vspace*{.4cm}
 \caption{
 Loading the dipole trap from the MOT with 100 \% efficiency. Shown are
 photon counts detected with the APD. Discrete signal levels correspond
 to an empty MOT (N=0, MOT stray light only), one and two atoms,
 respectively. Alternating to normal MOT operation the dipole trap
 is switched on and the MOT off for periods of 1 s. The signal decreases to
 the lowest level which is due to the dipole trap laser stray light only. After that
 the atoms are recaptured into the MOT showing the same fluorescence level as before.
 During normal MOT operation
 atoms are occasionally loaded into the MOT from the background vapor
 (L) or leave the MOT (C, in this particular case
 as a result of a cold collision [13])
 }
 \label{stufe1}
 \end{figure}

 The relevant part of the apparatus is shown in Fig.  \ref{setup}, details have
 been described elsewhere \cite{Haubrich 96,Gomer 98}. The number of trapped atoms
 and its temporal fluctuations in the MOT are determined by the balance between
 loading from the atomic vapor and different loss mechanisms removing atoms from
 the trap. For experiments with few atoms a strong reduction of the capture rate
 from the gas phase is essential. In the first place, this can be achieved by
 reducing the partial atomic vapor pressure in the vacuum chamber. Stiff magnetic
 field gradients lead to an additional reduction of the capture rate \cite{Hoepe 93}.
 The intratrap collision loss rate in our MOT with a gradient of $B'=dB/dz= 375$
 G/cm is comparable to the rate for background collisions \cite{Ueberholz 99}.
 Along with the base pressure below 10$^{-10}$ mbar this governs the trapped atom
 number dynamics on a time scale of seconds (see Fig.  \ref{stufe1}). The average
 number of trapped atoms can easily be adjusted between 1 and 10 by changing the
 Cesium vapor pressure.

 We emphasize that a strong magnetic field gradient is not a necessary condition
 for trapping of small atom numbers, as has been demonstrated previously
 \cite{Hu94}. However, a small MOT volume tremendously improves the localization
 of trapped atoms. The spatial distribution of the fluorescence in our
 MOT measured by a CCD camera has a Gaussian distribution with $1/e$ radius between
 5 and 17 $\mu$m depending on the laser intensity. The trap size was observed to be
 independent of the atom number $N$ up to $N=8$.

 An atom strongly driven by resonant laser light emits an average fluorescence power
 of $\hbar \omega \Gamma/2 \approx 3$ pW, where $\Gamma$ is the natural linewidth,
 $\Gamma=2\pi\times 5.2$ MHz for Cesium. With a realistic overall detection
 efficiency of 10$^{-3}$ it is necessary to discriminate about 3 fW fluorescence
 power from a stray light background. Fluorescence of the atoms trapped in the MOT
 is observed with an avalanche photodiode (APD) in single photon counting mode with
 a measured photon detection efficiency of 50 \% at $\lambda=852$ nm.
 The fluorescence light is collected by a lens mounted inside the vacuum chamber and
 is sent through a telescope onto the APD, see Fig. \ref{setup}. Typical photon
 counting rates are $3-20\cdot10^3$ s$^{-1}$ per atom depending on the detuning and
 intensity of the trapping laser. Well separated equidistant steps in the fluorescence
 signal allow us to monitor the number of trapped atoms in a non-invasive way and in
 real time \cite{Gomer 98}, see Fig. \ref{stufe1}. During normal
 MOT operation we can thus
 easily choose a desired atom number to be transferred into the dipole trap.

 The dipole trap consists of a single tightly focused Nd:YAG laser beam which is
 linearly polarized and superimposed on the MOT. We use the same lens inside the
 vacuum chamber both for focusing the dipole trap laser and for collecting the
 fluorescence. Due to the large difference between the wavelengths of the Nd:YAG
 laser (1064 nm) and the D$_2$ line of Cesium (852 nm), dipole trap laser radiation
 is easily blocked from the detection by interference filters.
 During simultaneous operation of both
 traps the fluorescence of the trapped atoms is substantially reduced due to the
 light shift. Thus, the optimal geometrical overlap of the dipole laser with the MOT
 trapping volume can be achieved by minimizing the fluorescence. The dipole trap
 laser has a waist of about 5 $\mu$m yielding a trap depth corresponding to 16 mK
 and a maximum photon scattering rate of 190 s$^{-1}$ at the center of the trap for
 a typical laser power of 2.5 W.

 Transfer of atoms between the two traps is accomplished through suitable timing
 sequences. First only the MOT is operated to collect atoms. To transfer these atoms
 into the dipole trap, the Nd:YAG laser is turned on a few ms before the MOT lasers
 are turned off. To recapture the atoms into
 the MOT, this procedure is reversed. By analyzing the resonance fluorescence we
 measure without any uncertainty the number of atoms right before transferring them
 into the dipole trap and directly after reloading the MOT, see Fig. \ref{stufe1}.
 Note that the dipole trap provides a conservative potential and can therefore not
 capture atoms from the background vapor. The probability of capturing an atom by the
 MOT during the detection immediately after reloading the MOT is less than 1\% and
 can be neglected.

 Atoms can be caught by the dipole trap only at places where the dipole potential
 exceeds the atomic kinetic energy $E_{kin}$, being of the order of $k_B T_D$
 ($T_D=\hbar \Gamma/2k_B = 125 \mu$K for Cs) yielding the geometric loading
 efficiency $P=1- (E_{kin}/U)^{w_0^2/r_0^2}$, where $U>E_{kin}$ is the dipole
 potential in the trap center, $w_0$ and $r_0$ are $1/e$ radii of the dipole trap
 and the MOT, respectively. Even for small MOT sizes used in our experiments
 (typically $r_0=10 \mu$m) $P$ is about 70 \%. However, during few ms of
 simultaneous operation of both traps, the MOT effectively cools the atoms down
 into the dipole potential. We have indeed found that 1 s after loading $N$ atoms
 from the MOT, the probability to find the same $N$ atoms in the dipole trap is
 more than 98\% for all $N$ up to 7. This is consistent with the measured dipole
 trap lifetime (see below) and 100\% loading efficiency.

\begin{figure}[hbt]
 \center{\epsfig{file=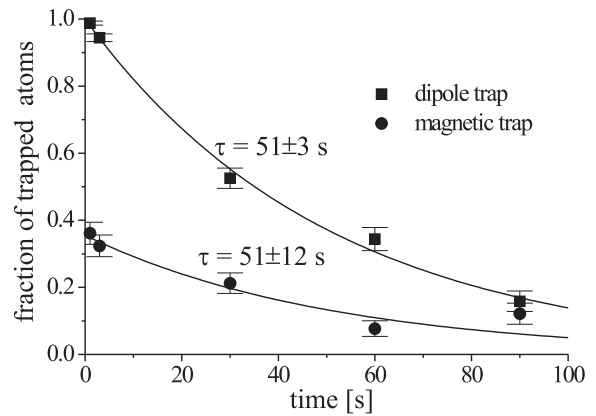,width=3in}}\vspace*{.4cm}
 \caption{
 Lifetime measurement for two types of traps. For details see text.
 }
 \label{lifetime}
 \end{figure}

 Varying the time spent by atoms in the dipole trap we have measured the fraction
 of atoms transferred back into the MOT and hence the lifetime in the dipole trap
 to be also independent of $N$. The results are shown in Fig. \ref{lifetime}
 demonstrating again 100\% loading efficiency and a storage time of $51\pm 3$ s.
 Each point shows an averaging over 400 atoms (about 100 single observations on $N$
 atoms, $N$ varying from 1 to 7). The same procedure repeated with the dipole trap
 laser blocked is also presented in Fig. \ref{lifetime} as circle symbols. In this
 case atoms are stored in the quadrupole magnetic field as reported previously
 \cite{Haubrich 96}. As expected, roughly half of the atoms are immediately lost
 after switching off the MOT lasers due to a statistical distribution of the spin
 orientations relative to the local magnetic field. From the fact that the atoms'
 lifetime in the magnetic trap is the same, we conclude that storage in both,
 physically very different traps is limited by background gas pressure only.

 Due to the large detuning of the dipole trap laser from atomic resonances the
 light shifts of both hyperfine ground states are nearly identical. We thus
 expect equal lifetimes in the dipole trap for atoms prepared in either
 hyperfine state. Still any initial preparation in a hyperfine state will be
 destroyed by off-resonant photon scattering in the dipole trap, yielding
 a relaxation of spin polarization. In the experiment, optical pumping
 is used to prepare an atom in a certain state. Introducing a delay of 8 ms between
 switching off the MOT repumping laser (resonant with the transition $F=3\to F'=4$)
 and the MOT cooling laser (in close resonance with the transition $ F=4 \to  F'=5$)
 provides a convenient way to prepare the atoms in either $F=3$ or $F=4$ state.

 For state-selective detection the dipole trap laser is switched off by means of
 a Pockels cell 50 $\mu$s before the detection laser is turned on. As a detection
 laser the MOT cooling laser is used, now tuned into exact resonance with the
 transition $F=4 \to F'=5$, while the MOT repumping laser remains blocked. Atoms
 in the $F=4$ state contribute to the fluorescence signal, while atoms in the
 $F=3$ state stay dark.

\begin{figure}[hbt]
 \center{\epsfig{file=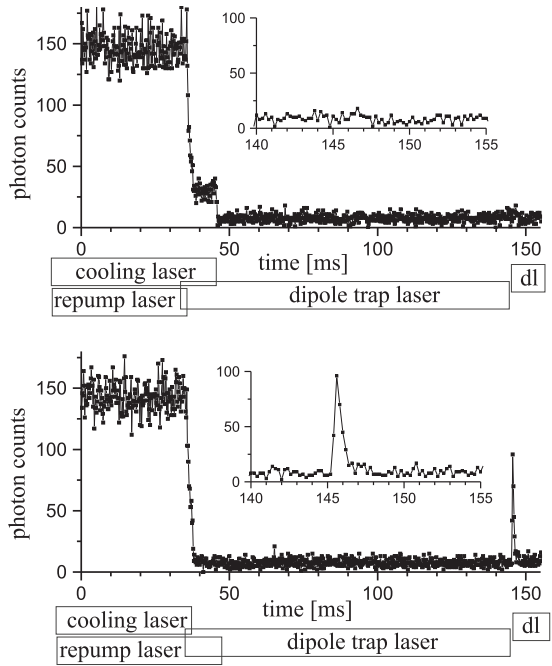,width=3in}}\vspace*{.4cm}
 \caption{
 State-selective detection of atoms in the $F=4$ state after 100 ms in the dipole trap.
 Integration time for one point is 200 $\mu$s. Frames below time axes show the switching
 sequence of all lasers used in the measurement and indicate their temporal overlaps,
 dl - detection laser. Top: atoms are prepared in the $F=3$ state. Bottom: after
 preparation in the $F=4$ state.
 }
 \label{detect}
 \end{figure}

\begin{figure}[hbt]
 \center{\epsfig{file=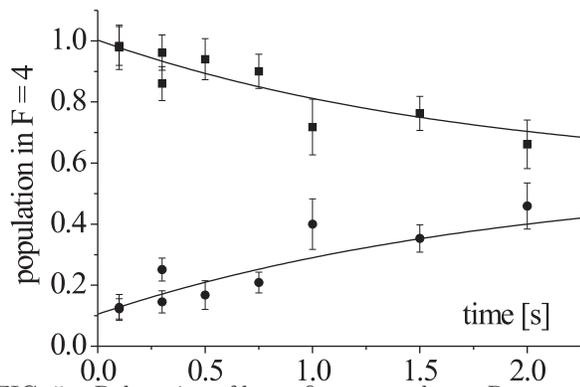,width=3in}}
 \caption{
 Relaxation of hyperfine states due to Raman scattering. Shown is the measured
 population in $F=4$ state after initial preparation in $F=4$ (squares) and in
 $F=3$ (circles).
 }
 \label{relaxplot}
 \end{figure}

 A typical measurement sequence is shown in Fig. \ref{detect}. After a storage
 of 100 ms in the dipole trap one clearly observes a fluorescence burst in the
 case that the $F=4$ state is prepared and probed whereas there is no additional
 signal from atoms initially prepared in the $F=3$ state. By varying the time
 between preparation of atoms in the dipole trap and probing we observed the
 relaxation of both hyperfine states towards an equilibrium due to spontaneous
 Raman scattering from the dipole trap laser, as given in Fig.  \ref{relaxplot}.
 Each of these measurements was obtained by averaging over 90 atoms (30 runs with
 3 trapped atoms each), following the scheme outlined above. The photon number in
 the detected fluorescence peak is proportional to both the population in the $F=4$
 state and the stored atom number, the latter known exactly. The figure shows the
 population in the $F=4$ state extracted from the obtained data. Theoretical
 treatment reveals that the Raman scattering rate transferring population from
 one hyperfine state $F$ to the other $F'$ is proportional to $2F'+1$.
 Spin-changing collisions can be neglected in this analysis \cite{HCC}.

 The observed relaxation rate is almost three orders of magnitude smaller than
 the expected total photon scattering rate at the trap center. The measured very long
 ground-state relaxation times of $4.2\pm 0.7$ s for $F=4$ and $3.3\pm 0.6$ s
 for $F=3$ clearly show a strong suppression of spontaneous Raman scattering
 processes. This effect is due to destructive interference of scattering
 amplitudes far from resonance and has recently been observed on optically
 trapped Rubidium atoms and explained in \cite{Cline 94}. For a Cs atom in a
 Nd:YAG dipole trap Raman scattering is suppressed compared to the Rayleigh
 scattering rate by a factor of 90 \cite{mf}. Although the experiment described
 here is only sensitive to the change of the hyperfine $F$ state, similarly long
 relaxation times are also expected for all Zeeman $m_F$-sublevels.

 The remaining factor of about 8 in the ratio between the measured relaxation
 rate and the maximum total scattering rate can be explained by the oscillatory
 motion of the atoms in the dipole trap. The scattering rate is proportional to
 the time-averaged light intensity seen by the trapped atoms. In a
 gaussian-shaped potential
 an average intensity 8 times smaller than the maximum intensity in the trap center
 corresponds to an oscillation amplitude of about 7 $\mu$m (1.5 $w_0$).
 This estimation is supported by the following simple model
 of the loading process: During simultaneous operation of both
 traps the MOT cooling action is still at work for large
 distances from the trap center (a distance of 1.5 $w_0$ corresponds
 to a light shift of 10 $\Gamma$ in our case). Thus the MOT damps the
 atomic motion and effectively collects the atom into the dipole trap.
 Near the dipole trap center, however,
 the MOT light forces are too weak to enforce a further localization
 and after switching off the
 MOT we expect the corresponding oscillation amplitude of about
 1.5 $w_0$ in the conservative dipole potential.

 As already mentioned, for state-selective detection we use the MOT cooling laser
 beams. As a result of the 3D-character of the MOT light field the atom decays
 within about 400 $\mu$s into another hyperfine state thus terminating the
 fluorescence signal. Even in this situation we are able to detect on average 3
 fluorescence photons per atom on a stray light background of 0.5 photons. There
 are natural ways to improve the detection scheme by using closed transitions in
 an optically pumped atom similar to single trapped ion experiments \cite{Itano 93}
 and setting up different laser beam geometries. This should result in an
 improvement of the signal to noise
 ratio enabling us to detect the state of a single trapped atom with high certainty.

 In summary we have demonstrated 100 \% transfer efficiency of atoms from the MOT
 into the dipole trap. For times short compared to both the dipole trap storage
 time and the spin relaxation time, the experiment also shows the potential of the
 apparatus to prescribe the state of a exactly determined number of atoms. It is
 now conceivable to tailor dipole trapping laser fields for transporting a desired
 number of atoms to another location (optical tweezer for single atoms), e.g. into
 a resonator of high finesse. Together with state-selective detection this will
 open up a wide range of applications not possible so far with either single
 trapped ions or many neutral atoms in optical dipole traps.

 Usual sources of neutral atoms like atomic beams or atoms released from a
 magneto-optical trap provide a flux of uncorrelated atoms. For some specific
 experiments,
 however, there is great interest in a source providing an arrival of a certain
 small number of atoms at time moments set by the experimentalist, i.e. with a
 $\delta$-like arrival probability distribution.

 A promising application are experiments on cavity QED and quantum information
 processing. Quantum logic gates can be implemented by entangling neutral atoms
 through the exchange of cavity photons. The feasibility of using this atom-cavity
 interaction in the optical range has already been demonstrated by the groups of
 J. Kimble  \cite{Kimble} and G. Rempe \cite{Rempe}. In particular they have trapped
 a single atom inside a cavity. However, in these experiments  the atoms enter the
 cavity in a random way after being released from a MOT and it
 is impossible to have a certain small number of atoms in the cavity on demand. As
 noted in \cite{Pellizari 95} simultaneous strong coupling of more than one atom to
 the cavity can minimize the effects of decoherence (cavity decay) in a quantum gate
 implementation. This would produce an entanglement between internal states of two
 atoms -- a basic element for quantum information processing.

\begin{thebibliography}{99}
\bibitem{HCC}
  Spin-changing collisions release a kinetic energy of at least
 0.2 K per atom for Cs which is much larger than the dipole trap depth and
 lead to immediate loss of trapped atoms. Since the measured lifetime is independent
 of the atom number, the collisional rate in the dipole trap is smaller than 0.02 s$^{-1}$.
\end{thebibliography}
\end{document}